\documentclass[%
 reprint,
 amsmath,amssymb,
 aps, prl,
]{revtex4-2}

\usepackage{morefloats}
\usepackage{color}
\usepackage{epsfig,graphicx,amsfonts,amsbsy}
\usepackage{amsmath,amsfonts,amsthm,amssymb}
\usepackage{appendix}
\usepackage{bbm}
\usepackage{makeidx}
\usepackage{url}
\usepackage{verbatim}
\usepackage{mathrsfs} 
\usepackage{morefloats}
\usepackage{appendix}
\usepackage{bbm}
\usepackage{makeidx}
\usepackage{url}
\usepackage{verbatim}
\usepackage[bookmarksnumbered,pdfpagelabels=true,plainpages=false,colorlinks=true,linkcolor=blue,citecolor=blue,urlcolor=blue]{hyperref}
\usepackage[bookmarksnumbered,pdfpagelabels=true,plainpages=false,colorlinks=true,linkcolor=blue,citecolor=blue,urlcolor=blue]{hyperref}
\usepackage{array}
\usepackage{booktabs}
\usepackage{multirow}
\usepackage{bbm}
\usepackage{tabularx}
\usepackage{cancel,soul}
\usepackage{nicefrac, xfrac}
\usepackage{ulem}
\usepackage{bbold}

\newcommand{\bs}{\boldsymbol{{\rm s}}}
\newcommand{\bh}{\boldsymbol{{\rm h}}}

\newcommand{\ba}{\boldsymbol{{\rm a}}}

\newcommand{\bH}{\boldsymbol{\rm H}} 
\newcommand{\bE}{\boldsymbol{\rm E}}

\newcommand{\bgamma}{\boldsymbol{{\rm \gamma}}}

\newcommand{\bbeta}{\boldsymbol{{\rm \beta}}}

\newcommand{\bepsilon}{\boldsymbol{{\rm \varepsilon}}}

\newcommand{\br}{\boldsymbol{{\rm R}}}
\newcommand{\bmm}{\boldsymbol{{\rm m}}}
\newcommand{\bn}{\boldsymbol{{\rm n}}}

\newcommand{\bq}{\boldsymbol{{\rm q}}}
\newcommand{\bP}{\boldsymbol{{\rm P}}}
\newcommand{\bp}{\boldsymbol{{\rm p}}}

\newcommand{\bg}{{{\rm g}}}
\newcommand{\bz}{\boldsymbol{{\rm z}}}

 \newcommand{\bff}{\boldsymbol{{\rm f}}}
\newcommand{\cL}{{{\cal L}}}
\newcommand{\cF}{{{\cal F}}}
\newcommand{\cA}{{{\cal A}}}
\newcommand{\cT}{{{\cal T}}}
\newcommand{\cP}{{{\cal P}}}
\newcommand{\cS}{{{\cal S}}}
\newcommand{\cK}{{{\cal K}}}
\newcommand{\balpha}{\boldsymbol{{\rm \alpha}}}

\newcommand{\sectioncustom}[1]{\noindent {\it #1.- }}

\usepackage{xcolor}
\usepackage{nicefrac, xfrac}
\def \usach {Departamento de F\'isica, Universidad de Santiago de Chile, 9170124, Santiago, Chile.}
\def \cedenna {Centro  de Nanociencia y Nanotecnología CEDENNA, Avda. Ecuador 3493, Santiago, Chile.}
\def \fcfm {Departamento de F\'isica, FCFM, Universidad de Chile, Santiago, Chile.}

\begin{document}

\preprint{APS/123-QED}

\title{Elementary theory of Magnetoferrons: bringing magnons and ferrons together in multiferroic systems}

\author{Mario A. Castro$^{1}$}
\author{Carlos Saji$^{1}$}
\author{Guidobeth Saez$^{1}$}
\author{Patricio Vergara$^{1}$}
\author{Sebastian Allende$^{2,3}$}
\author{Alvaro S. Nunez$^{1}$}

\affiliation{${}^{1}$\fcfm}
\affiliation{${}^{2}$\usach}
\affiliation{${}^{3}$\cedenna}



\date{\today}

\begin{abstract}
The collective excitations of a multiferroic material are analyzed. We show that these excitations also exhibit magnetoelectric behavior, leading to the hybridization of magnons ,oscillations of the magnetization field, and ferrons, which are oscillations of the electric dipolar density field. We term these emergent entities 'magnetoferrons', study their main properties, and discuss their potential applications. Additionally, we provide a phenomenological framework for these systems, which will be invaluable for describing the dynamics of the multiferromagnetic state.
\end{abstract}

\maketitle

\sectioncustom{Introduction}
The multiferromagnetic phases of matter present excellent prospects for realizing several novel physical effects that could be pivotal for new technological developments\cite{Spaldin2019, Song2022, Bennett2024, Castro2024}. One such effect is the magnetoelectric effect\cite{Mostovoy2024}, in which electric stimuli produce magnetic responses and vice versa\cite{Kimura2003}. Such a coupling is often regarded as a signature of profound microscopic mechanisms dictating quantum mechanical electronic behavior and, therefore, has received much attention. In this work, we analyze the collective excitations of a multiferroic material and show that they, too, display this magnetoelectric behavior that results in a hybridization of the magnons, oscillations of the magnetization field, and the ferrons, oscillations of the electric dipolar density field. We call these emergent entities magnetoferrons, study their main properties, and discuss their potential applications.

The appearance of massless or gapless Goldstone modes characterizes the phenomenon of spontaneous breaking of continuous symmetries. These modes are related to slow, symmetry-restoring, long-wavelength fluctuations in the order parameter.  In crystals, for instance, the vibrational modes known as acoustic phonons\cite{Gan2019} are connected to slow fluctuations in the density of the crystal's atoms. 
Magnons are quantum particles corresponding to Goldstone modes associated with the rotational symmetry destroyed by the ferromagnetic order\cite{Anderson2018}. They are elementary excitations with a gapless dispersion relation $\omega\sim k^2$, which transport momentum, energy, and magnetization currents. Magnonics\cite{Chumak2015, Pirro2021, Yuan2022, Flebus2024, Harms2024, Saji2023, Jiang2023, Aguilera2020, Bunkov2020} is an emerging field of research that focuses on the study and manipulation of magnons. These spin waves can carry and process information, making them promising candidates for developing novel, energy-efficient devices and technologies for information processing and communication.

The field of magnonics explores ways to generate, detect, and manipulate magnons and their associated spin waves. The purpose of this endeavor is to understand the fundamental physics of magnonic phenomena and to develop practical applications, such as magnonic circuits\cite{Chumak2014, Cramer2018, Wu2018, Cornelissen2018}, logic devices\cite{Kostylev2005, Ganzhorn2016}, and Josephson interferometers\cite{Troncoso2011, Troncoso2014, Nakata2014, Nakata2024}, which can operate at high speeds with low power consumption. 

Ferrons, theoretically proposed as quantum particles related to the elementary excitations of electric polarization in ferroelectric materials, do not follow this pattern. For ferroelectric, these particles follow a dispersion $k \sim \omega^2 \epsilon(\omega)$, where $\epsilon$ corresponds to the dynamic permittivity \cite{Born1955,Tang2022}.
Bosonic excitations in ferroelectrics that carry electric dipoles are identified from the phenomenological Landau-Ginzburg-Devonshire theory\cite{Levanyuk2020}. Ferrons are shown to exist from the simultaneous action of anharmonicity and broken inversion symmetry. Unlike magnon excitations, which are transverse perturbations of the magnetic order,  ferron quasiparticles in displacive ferroelectrics are of the longitudinal kind. Based on the predicted ferron spectrum, temperature-dependent pyroelectric and electrocaloric properties, electric-field-tunable heat and polarization transport, and ferron-photon hybridization are predicted in \cite{Tang2022}.

In multiferroics\cite{Spaldin2010, Wang2016, Spaldin2019, Song2022}, magnons and ferrons naturally hybridize, giving rise to magnetoferrons.
Magnetoferrons might become handy in several contexts, such as when exploiting the sensitivity of multiferroic materials to electric and magnetic fields. In this sense, they can be used to develop susceptible sensors and actuators that respond to multiple stimuli. Magnetoferrons are different from electromagnons\cite{Kimura2014, Kubacka2014}, i.e., magnon excitation by oscillatory electric fields of electromagnetic wave perturbations.

From a spintronics point of view, storage devices based on multiferroic components can be envisioned. The dual magnetoelectric properties of multiferroic materials can be leveraged to develop nonvolatile data storage devices where information is encoded in electric polarization and magnetization. In this context, magnetoferrons could provide low-dissipation mechanisms to retrieve written information nondestructively.
Finally, magnetoelectric waves can be used to design new types of devices for signal transmission and processing, especially in the microwave frequency range.

\sectioncustom{Phenomenological theory}
Our selected model is based on previous work \cite{Saez2023, Vergara2024, Saez2024}, where a novel mechanism for multiferroicity was proposed relying on a well-established time reversal ($\cT$) breaking antiferromagnetic model. This model was enhanced by intentionally breaking the space inversion ($\cP$) while keeping the symmetry $\cP\cT$. Such symmetry breaking gave rise to magnetoelectric features readily interpreted as multiferroic behavior. In this section, we will provide a phenomenological account of those systems that will prove invaluable in addressing the dynamic description of the multiferroic state.

In the Landau theory of ferroelectrics, symmetry requirements are crucial for determining the system's possible phases and properties. 
Ferroelectrics must break the inversion symmetry to exhibit spontaneous polarization. This means that the crystal structure of the ferroelectric phase lacks a symmetry center.
The order parameter in Landau theory is usually the polarization vector $\mathbf{P}$. The free energy expansion must be invariant under the symmetry operations of the high-symmetry (paraelectric) phase.
 Given that the free energy of the ferroelectric part can be written as $\cF_e(\bP)=\left(\frac{\cA}{2} (\partial_i \bP)^2+\frac{\balpha}{2}\bP^2+\frac{\bbeta}{4}\bP^4+\frac{\bgamma}{6}\bP^6\right)-\bE\cdot\bP$. Here, $\balpha$, $\bbeta$, and $\bgamma$ represent the Landau coefficients that will be regarded as positive. This locates our isolated electric system in the paraelectric phase.
In external fields (electric, stress, etc.), additional coupling terms in the free energy expansion can express the Landau-Ginzburg-Devonshire Lagrangian expansion\cite{rabe2007}.

The Landau theory of ferroelectrics requires that the free energy expansion and all related terms respect the symmetry of the paraelectric phase. This includes considerations of inversion symmetry breaking, proper treatment of the order parameter, and appropriate coupling terms for external fields and mechanical strain. These symmetry requirements are essential for accurately describing ferroelectric materials' phase transitions and properties.

In the Landau theory of antiferromagnets, the description includes both the magnetization field $\mathbf{m}$ and the Néel field $\mathbf{n}$. 
The order parameters in antiferromagnets are the magnetization field $\mathbf{m}$ and the Néel field $\mathbf{n}$. The Néel field $\mathbf{n}$ represents the staggered magnetization, indicating the difference in magnetization between sublattices, while the magnetization field $\mathbf{m}$ represents the net magnetic moment per unit volume.

\begin{figure}
    \centering
    \includegraphics[width=0.99\linewidth]{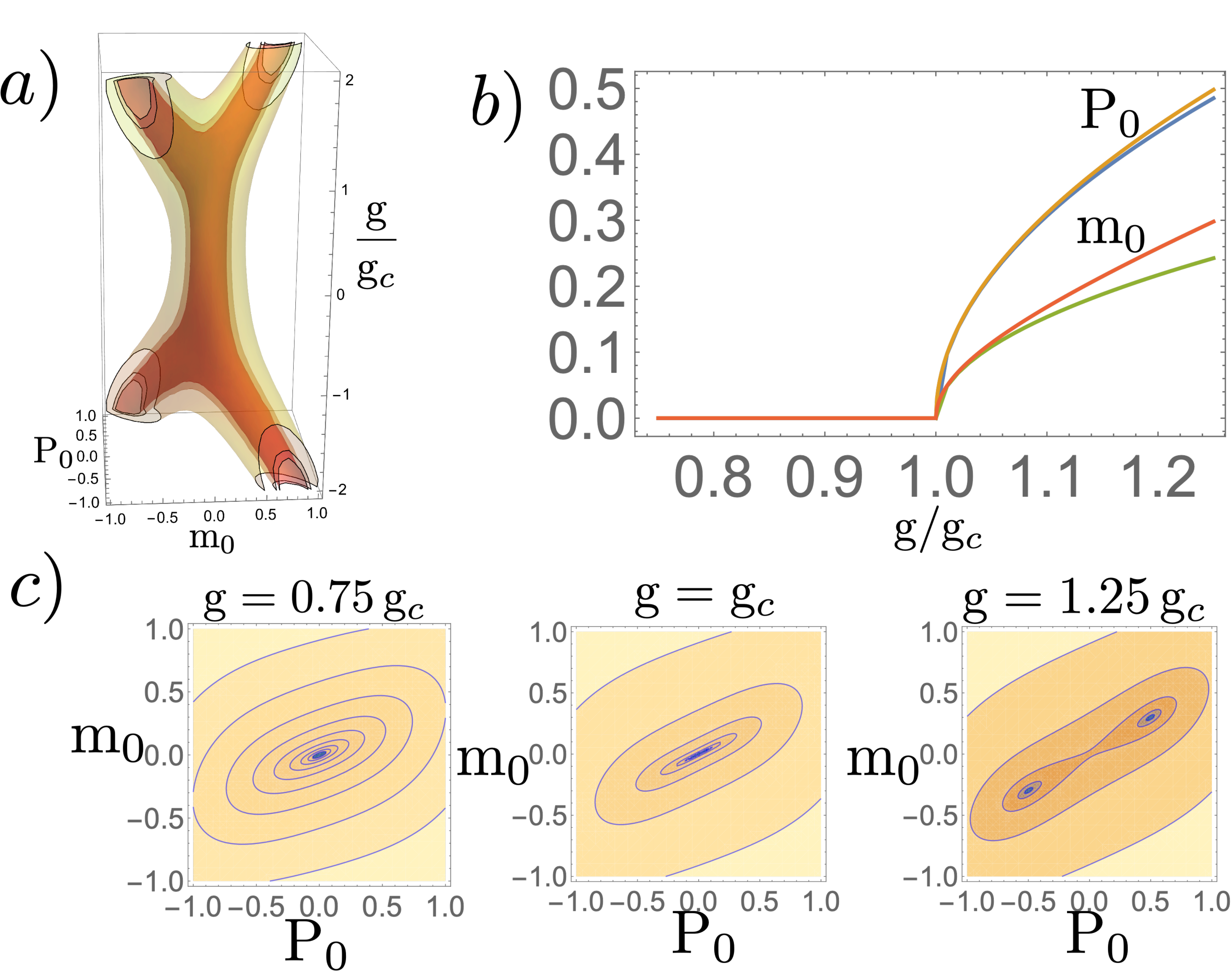}
    \caption{Stability analysis of the multiferroic state in terms of the coupling parameter $\bg$. $a)$ Energy levels, in the $(m,p,\bg)-$ space of the functional energy $\cF(m,p,\bg)-\cF_{\rm min}(\bg)$. It vividly illustrates the nature of the bifurcation and the origin of the multiferroic state. $b)$ Zoom at the bifurcation displaying the pitchfork-like behavior for both polarization and magnetization. Along with the exact numerical results, we attach the results of the Landau analysis that are in great agreement, $c)$ Snapshots of the energy levels in $(m,p)-$ space at different values of $\bg$, $\bg<\bg_c$, $\bg=\bg_c$, and $\bg>\bg_c$. They illustrate how the transition unfolds.  }
    \label{fig: multiferroic}
\end{figure}

Just like for the ferroelectric part, symmetry plays a crucial role in the Landau theory of antiferromagnet degrees of freedom. The antiferromagnet order parameter maintains the inversion symmetry, which means that the order parameter $\mathbf{n}$ changes sign under inversion, while $\mathbf{m}$ remains unchanged. This inversion symmetry is essential to define the free energy expansion and the interactions between the order parameters and the external fields\cite{Auerbach1994, Borejsza2004, Sachdev2011}. The free energy of the antiferromagnetic system can be customarily expressed, up to the lowest order in the order parameter fields, as $\cF_m(\bn,\bmm)=\left(\frac{a}{2} \bmm^2+\frac{b}{4} \bmm^4 +\frac{A}{2} (\partial_i \bn)^2-\frac{K}{2}(\bn\cdot\bz)^2 - \bH\cdot\bmm\right)$, where $a>0$ and $b>0$ are Landau parameters, $A$ is a Giznburg-like parameter, $K$ and anisotropy related constant, and $\bH$ and external magnetic field. The Lagrangian $\cL$ is expanded as a function of both $\mathbf{m}$ and $\mathbf{n}$, ensuring that all terms respect the symmetry of the crystal\cite{Rezende2019, ElKanj2023, Huang2024}. 
The collective Lagrangian density becomes:
\begin{align}
    \cL&=\frac{\rho}{2}\dot{\bP}^2-\cF_e(\bP)+s\,\bmm\cdot\bn\times\dot{\bn}-\cF_m(\bn,\bmm)-\bg\, \bP\cdot\bmm 
\end{align}
where $s=S/\gamma$, is the spin density over the gyromagnetic ratio.
The last term, proportional to $\bg$, breaks the system's time-reversal invariance ($\cT$) while simultaneously breaking its inversion-reversal symmetry $\cP$.
However, the coupling is $\cP\cT$ invariant. In this sense, it is related to the axion electrodynamics in topological materials\cite{Sekine2021}. It constitutes a coupling between the ferroelectric and magnetic degrees of freedom, as shown in Fig. (\ref{fig: multiferroic}). When large enough, this coupling introduces a second-order symmetry-breaking transition to a multiferroic state, showing ferroelectricity and magnetism\cite{Kamenetskii2023}.
Needless to say, the specific details of our calculations do not depend on the specific form of this coupling; similar calculations can be achieved for a coupling of the more familiar form $\sim (\bmm\cdot\bP)^2$\cite{Wang2016}.

The equations of motion can be written as:
\begin{eqnarray}
s\dot{\bn}&=&(\bff_{\bmm} - \bg\; \bP)\times \bn\\   
s\dot{\bmm}&=&\bff_{\bn} \times \bn+\bff_{\bmm} \times \bmm\nonumber\\
\rho\ddot{\bP}&=&\bff_{\bP}-\bg\, \bmm, \nonumber
\end{eqnarray}
where $\bff_{\bh}=-\delta \cF_m(\bn,\bmm)/\delta\bh$ and $\bff_{\bP}=-\delta \cF_e(\bP)/\delta \bP$. These equations constitute the basic mathematical framework for describing the basic magnetoelectric phenomena to be addressed in the remaining work.


\sectioncustom{Equilibrium conditions}
The energy minimum is attained for $\bn_0=\bz$. Depending on the value of $\bg$ we have $\bmm=\bP=0$ for $|\bg|<\bg_c$ and $\bmm={\rm m}_0 \hat{\bmm}$ and $\bP={\rm P}_0 \hat{\bP}$ for $|\bg|>\bg_c$. In the latter case $\hat{\bmm}={\rm sign}({\bg})\hat{\bP}$. This transition corresponds to a phase transition from an antiferromagnetic state towards a multiferroic phase.
An analysis of the equilibrium conditions, $\bg{\rm m}_0=\alpha{\rm P}_0+\beta{\rm P}^2_0$ and $\bg {\rm P}_0=a{\rm m}_0+b{\rm m}^2_0$, in the vicinity of the transition leads to a critical value $\bg_c=\sqrt{a\alpha}$. Furthermore, we found that the transition to the multiferroic state is a second-order one. We obtain for $\bg\gtrsim \bg_c$: ${\rm P}_0= \sqrt{\frac{ 2 a \bg_c }{a^2
   \beta +\alpha ^2 b}}
   \left(\bg-\bg_c\right)^{\sfrac{1}{2}}$ and ${\rm m}_0=\sqrt{\frac{ 2 \alpha \bg_c }{a^2
   \beta +\alpha ^2 b}}
   \left(\bg-\bg_c\right)^{\sfrac{1}{2}}$.

We add coupling of the electromagnetic field to the energy of the system with $\bE\parallel\bH\parallel\hat{\bmm}$. This leads to a contribution of the form $\delta\cF=-{\rm E}{\rm P}_0-{\rm B}{\rm m}_0$, the susceptibilities of the system can be readily written as:

\begin{eqnarray}
\chi_{{\rm e}}&=&\left(\frac{\partial{\rm P}_0}{\partial {\rm E}} \right)=\frac{ \cS(\sfrac{\bg}{\bg_c})}{2\alpha|\sfrac{\bg}{\bg_c}-1|}\label{eq: Susceptibilities} \\
   \chi_{{\rm m}}&=&\left(\frac{\partial{\rm m}_0}{\partial {\rm B}} \right)=\frac{\cS(\sfrac{\bg}{\bg_c})}{2 a|\sfrac{\bg}{\bg_c}-1|}\nonumber
\\
\chi_{{\rm me}}&=&\left(\frac{\partial{\rm P}_0}{\partial {\rm B}} \right)=\left(\frac{\partial{\rm m}_0}{\partial {\rm E}} \right)=\frac{\cS(\sfrac{\bg}{\bg_c})}{2 \bg_c|\sfrac{\bg}{\bg_c}-1|},\nonumber
\end{eqnarray}
donde $\cS(x)=
\begin{cases}
			1, & \text{if $x<1$ }\\
         \frac{1}{2}, & \text{if $x>1$ }
    \end{cases}.$ We note that $\chi_{{\rm me}}=\sqrt{\chi_{{\rm e}}\chi_{{\rm m}}}$. This is the maximum allowable magneto-electric susceptibility. This bound is achieved through thermodynamical stability requirements\cite{ODell1963}. The multiferroic transition displays a divergent magnetoelectric susceptibility. 
    
    A significant magnetoelectric response, where electric and magnetic fields strongly interact, has several key applications in advanced technologies. It enables the development of more efficient data storage devices, such as magnetoelectric random access memory\cite{Kosub2017}, sensitive sensors for medical and industrial use, and precise actuators for micromachines. In telecommunications, it facilitates the manipulation of compact and energy-efficient RF signals, while in spintronics, it allows faster and more efficient control of electronic devices by using electric fields to manipulate magnetic states\cite{Fusil2014}. In addition, magnetoelectric materials can convert mechanical energy into electrical energy, making them valuable for energy harvesting technologies\cite{SurezRodrguez2024}.
\begin{figure}[t!]
    \centering
    \includegraphics[width=0.99\linewidth]{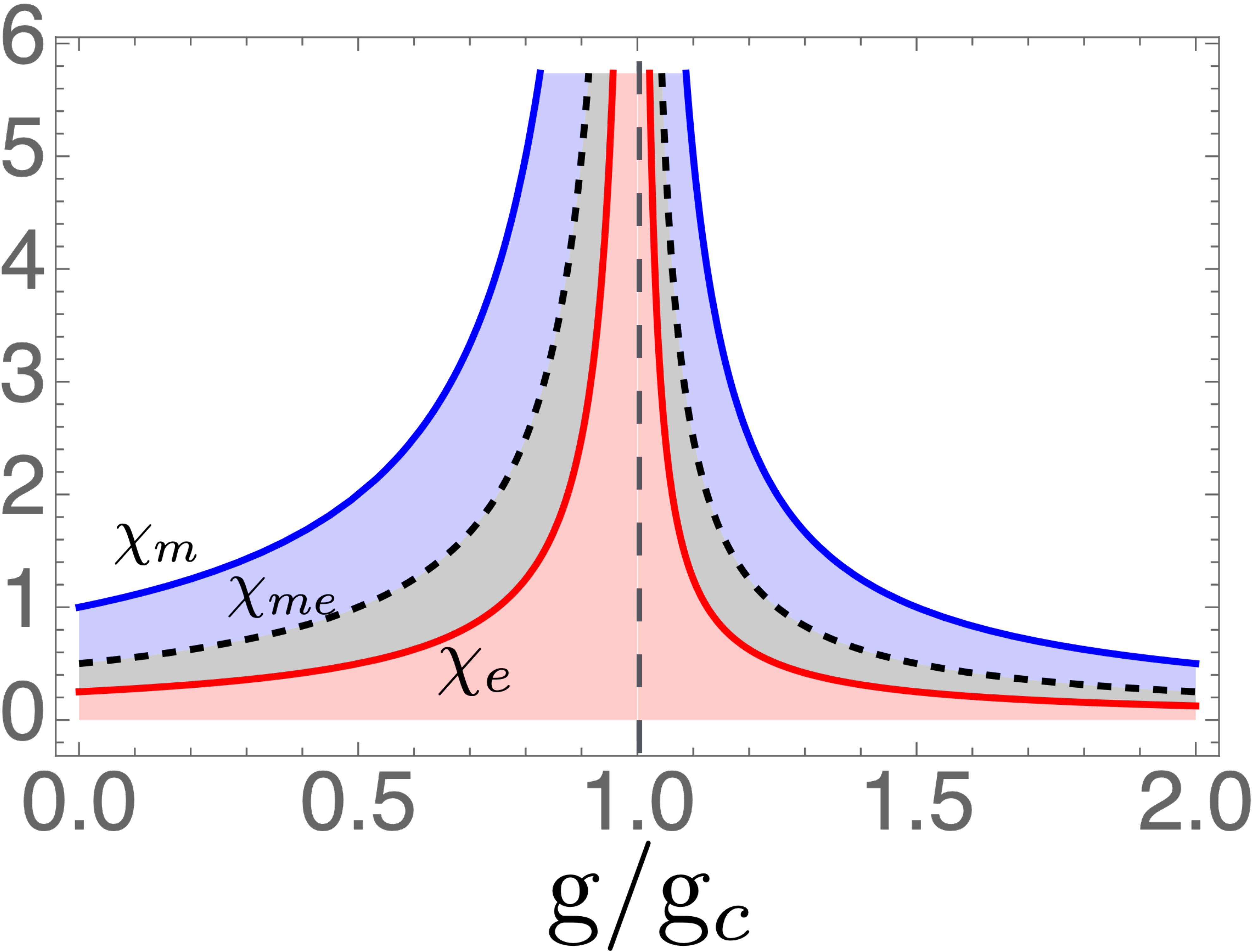}
    \caption{Susceptibilities from Eq. (\ref{eq: Susceptibilities}) as a function of $\bg$. The chosen parameters are $\alpha=2$ and $a=1/2$. The magnetoelectric susceptibility, $\chi_{me}$, is the geometric mean of $\chi_e$ and $\chi_m$. This corresponds to the highest magnetoelectric effect compliant with thermodynamic requirements.}
    \label{fig:enter-label}
\end{figure}
    
\sectioncustom{Spin wave spectrum} We expand the action around the equilibrium state  $\bmm={\rm m}_0 \hat{\bmm}+\delta\bmm$ and $\bP={\rm P}_0 \hat{\bP}+\delta \bp$.  We can write an effective action for the system with the $\delta\bmm$ fluctuations integrated out:
\begin{align}
    \cL^{(2)}&=\frac{\rho}{2}\delta\dot{\bp}^2-\frac{\cA}{2}(\nabla \delta\bp)^2+\frac{1}{2}\cK \delta \bp^2\\
    &+\frac{\chi_\perp}{2}\left(\left(s\,\bn\times\dot{\bn}+\bff-\frac{\bmm_0}{\chi_\perp}\right)^2-(\bff\cdot\bn)^2\right)\nonumber\\
    &+s\,\bmm_0\cdot(\bn\times\dot{\bn})-\frac{A}{2}(\nabla\bn)^2-\frac{K}{2}(\bn\cdot\bz)^2\nonumber
\end{align}
where we have defined $\bff=(\bH-\bg \delta\bp+\frac{\bmm_0}{\chi_\perp})$ and $\cK=(\alpha+3\beta\bP^2_0)$, $\chi_\perp=(a+3b\bmm^2_0)^{-1}$. This leads to the following equations of motion:
\begin{widetext}
\begin{align}
    \rho \ddot{\delta \mathbf{p}} &= \mathcal{A} \nabla^2 \delta \mathbf{p}-\mathcal{K}\delta \mathbf{p}  -\chi_{\perp}\bg (\mathbf{f}+s\mathbf{n}\times \dot{\mathbf{n}}- \dfrac{\mathbf{m}_0}{\chi_{\perp}}) +\chi_{\perp} \bg (\mathbf{f}\cdot\mathbf{n})\mathbf{n}\\
    \chi_{\perp} s^2  \mathbf{n}\times\ddot{\mathbf{n}} &= 2\chi_{\perp} s \mathbf{n}\times(\dot{\mathbf{n}}\times \mathbf{f} )  
    - \chi_{\perp} s \mathbf{n}\times(  \dot{\mathbf{f}} \times \mathbf{n})+A \mathbf{n}\times\nabla^2 \mathbf{n} + K (\mathbf{n}\cdot\mathbf{z})\mathbf{n}\times\mathbf{z}- \chi_{\perp} (\mathbf{f}\cdot \mathbf{n})\mathbf{n}\times\mathbf{f}\nonumber
\end{align}
\end{widetext}
The solutions to the wave equation can be expressed as $
\delta \bn(x,t) = e^{i \mathbf{k} \cdot \mathbf{x} - \omega t} \left(\bepsilon_1 n_1 + \bepsilon_2 n_2 \right).
$
The equation of motion unfolds into two coupled equations for \( n_1 \) and \( n_2 \) that can only be fulfilled by choosing a phase shift \(\pm \pi/2\) between them. The waves are, therefore, circularly polarized. Due to the magnetic field, there is a splitting between the two circular polarizations. We have,  
\[
\delta \dot{\bn} \times \bn_0 = \sigma \,\omega\, \delta \bn,
\] 
with $\sigma=+1 (-1)$ for right-polarized (left-polarized) waves. In the paraelectric regime, the equations for small deviations from equilibrium, $\bn_0=\bz$ and $\mathbf{H} = H_z \mathbf{z}$, are:

\begin{align}
    -\rho \omega^2 \delta \mathbf{p}_\perp &-=-\mathcal{A} k^2  \delta \mathbf{p}_\perp-\mathcal{K}\delta \mathbf{p}_\perp  +\chi_{\perp}\bg \left(\bg \delta \mathbf{p}_\perp\right.\nonumber\\
    &\left.+s\sigma \omega \delta \mathbf{n}_\perp\right)+\bg H_z \chi_{\perp} \delta \mathbf{n}_\perp\label{eq: eigensystem} \\
   - \rho \omega^2 \delta \mathbf{p}_z &=-\mathcal{A} k^2 \delta \mathbf{p}_z-\mathcal{K}\delta \mathbf{p}_z   + \bg \mathbf{m}_0\cdot \delta\mathbf{n}_\perp\nonumber\\
   -\omega^2 s^2  \chi_{\perp}\delta\mathbf{n}_\perp &=  2s \sigma \omega H_z  \chi_{\perp}  \delta \mathbf{n}_\perp +\bg\chi_{\perp}s \sigma \omega \delta \mathbf{p}_\perp-K \delta \mathbf{n}_\perp\nonumber\\
     &-A k^2 \delta \mathbf{n}_\perp-\mathbf{m}_0\left[\dfrac{1}{\chi_{\perp}} \mathbf{m}_0\cdot \delta \mathbf{n}_\perp -\bg \delta \mathbf{p}_z\right]\nonumber\\
     &+H_z^2 \chi_{\perp} \delta \mathbf{n}_\perp+ \bg H_z \chi_{\perp}\delta \mathbf{p}_\perp,
\end{align}
where the external field induces displacements in the ground state magnetic configuration: $\bmm^{eq}_0=\bmm_0+ \Delta\bmm$ and $\bn^{eq}_0= \bn_0+\Delta \bn$. In the linear field regime, these displacements can be approximated as $\Delta\bmm =  \dfrac{H_z \mathbf{m}_0^2}{ K} \hat{z}$ and $\Delta \bn = -\dfrac{H_z \bmm_0}{K} $. Additionally, the electric polarization experiences a displacement given by $\Delta\bP = -\dfrac{g \chi_{\perp} H_z^2 \mathbf{m}_0}{K(\mathcal{K} - \bg^2 \chi_{\perp})}+\mathcal{O}(H_z^3)$.

\begin{figure}
    \centering
    \includegraphics[width=1.0\linewidth]{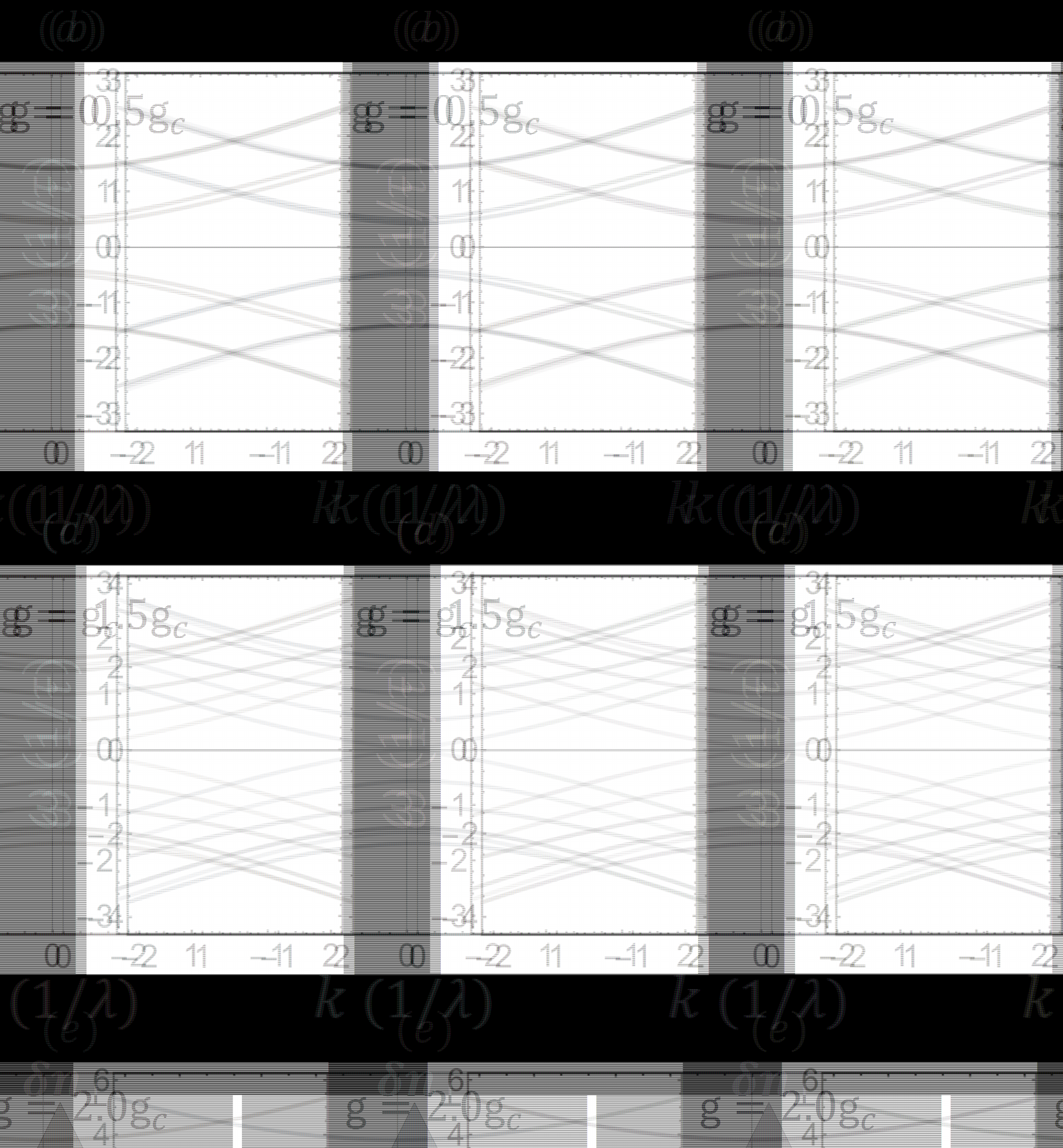}
    \caption{ Magnetoferrons dispersion relation for $a =1/2 $, $\alpha = 2$, $K = 0.5$, $H_z = 0$ and differents values of the magnetoelectric coupling constant $\bg$. The remaining parameters are taken as unity. The color of each band indicates the relative contributions of the components $\delta \mathbf{n}_\perp$ (red), $\delta \mathbf{p}_z$ (green) and $\delta \mathbf{p}_{\perp}$ (light yellow), as illustrated in the triangular color map. For $\bg = 0$, the bands shown in figure (a) are not hybridized. The magnetoferrons are independent for the electric and magnetic parts. For low coupling (b),  the degeneracy of the $\delta\mathbf{p}$-bands is lifted and slight hybridization is observed. However, the middle bands (green), associated with $\delta \mathbf{p}_z$ remain non-hybridized since $\mathbf{m}_0 = 0$. For $\bg = \bg_c$ (c), a multiferroic phase transition occurs, characterized by a vanishing gap. In this limit, we observe that the bands bands exhibit significant hybridization near $k = 0$. The hybridization of the bands is also observed for  (d) $\bg = 1.5 \bg_c$  and (e) $\bg = 2 \bg_c$.}
    \label{fig: bands}
\end{figure}

In the limit $\bP_0=\delta\bp=\bmm_0=0$, the equations reduce to standard antiferromagnetic dynamics\cite{Baltz2018, Troncoso2015, Rezende2020}. For $|\mathbf{m}_0|<<1$, two distinct branches emerge.  In the first branch $\delta\bp$ aligns with $\bn_0$ and decouples from the magnons, exhibiting a dispersion relation
$\rho\omega^2=\cA k^2+\cK$, see green line in Figs. (\ref{fig: bands}).(a-c). The second branch corresponds to $\delta\bp$ forming vector waves that share its circular polarization in the plane perpendicular to $\bn_0$. The spectrum can be readily found
\begin{align}
     &-\bg^2 \mathbf{m}_{0}^2 P+(P -\bg^2 \chi_{\perp} )\left(P[M + H_z\chi_{\perp} (H_z + 2 \sigma s\omega)]\right.\nonumber\\
     &\left. -\bg^2\chi_{\perp}^2( H_z  + s  \sigma \omega)^2\right) = 0
 \end{align}
where $P =\rho \omega^2 - \mathcal{A} k^2 - \mathcal{K}+\chi_{\perp} \bg^2$ and $M = -A k^2 - K - \dfrac{\mathbf{m}_0^2}{\chi_{\perp}}$. 

\begin{figure}
    \centering
    \includegraphics[width=1.0\linewidth]{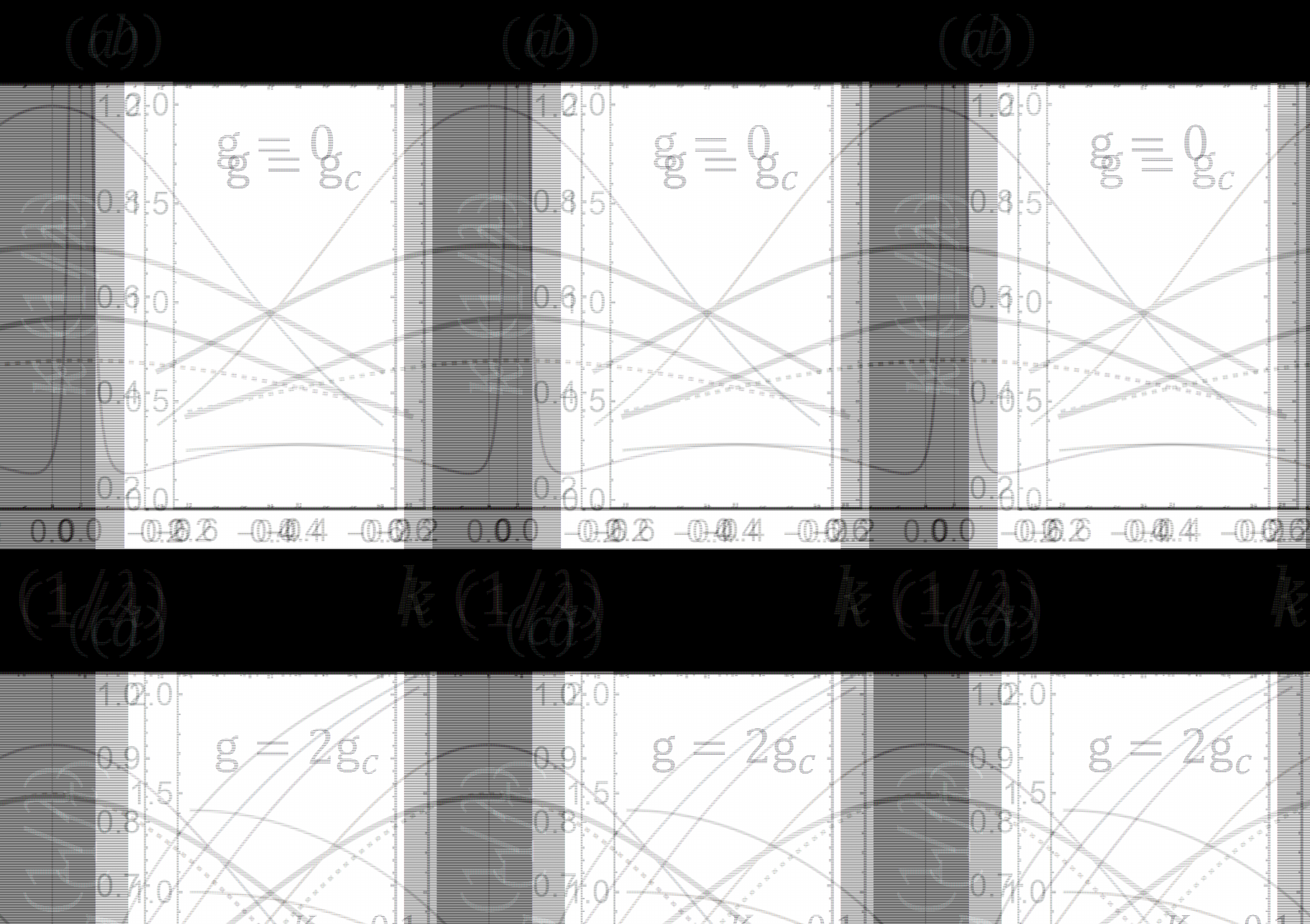}
    \caption{Panels (a), (b), and (c) show the curvature of the magnetoferron dispersion relation for the three lowest-frequency bands. A divergence in the curvature is observed at the multiferroic phase transition ($\bg = \bg_c$). (d) show the frequency gap of the magnetoferrons dispersion as a function of the magnetoelectric constant for three different values of magnetic anisotropy $K = 0.1$, $K = 0.5$, and $K = 1.0$. We consider that $a =1/2 $, $\alpha = 2$, $H_z = 0$, and the remaining parameters are taken as unity. }
    \label{fig: curvatures}
\end{figure}

The main features of the dispersion relations are depicted in Figs.(\ref{fig: bands}) and (\ref{fig: curvatures}). 
The multiferroic phase transition ($\bg=\bg_c$) is related to a vanishing of the gap as can be seen in Figs.(\ref{fig: bands}c) and (\ref{fig: curvatures}d).

The above account allows us to use a magnetoferron-based device specially tuned to sense electric or magnetic fields. The critical element of such a device is a high-Q cavity with the magnetoelectric material as an active substance. Magnetoferrons will be excited at specific resonances that will sharply depend on the values of the electric and magnetic fields.

\sectioncustom{Conclusion and outlook}
The collective excitations of a multiferroic material were analyzed. We showed that they, too, display this magnetoelectric behavior that results in a hybridization of the magnons, oscillations of the magnetization field, and the ferrons, oscillating the electric dipolar density field. We call these emergent entities magnetoferrons and discuss their potential applications. We also provided a phenomenological account of those systems that will prove invaluable in addressing the dynamic description of the multiferroic state.
In principle, these excitations are quantized in the form:
$$
\delta \bs_\ell(\br)=\sum_{\bq,\nu} \left({\rm t}_{\bq, \ell, \nu}\ba^{\phantom \dagger}_{\bq,\nu}\frac{{\rm e}^{i\bq\cdot\br}}{\sqrt{\omega_{q,\nu}}}+{\rm t}^*_{\bq, \ell, \nu}\ba^{\dagger}_{\bq,\nu}\frac{{\rm e}^{-i\bq\cdot\br}}{\sqrt{\omega_{q,\nu}}}\right),
$$
where $\delta\bs_\ell(\br)=(\delta\bn_\perp,\delta\bp_\perp,\delta\bp_z)$,   $\ba^{\dagger}_{\bq,\nu}$ ($\ba^{\phantom \dagger}_{\bq,\nu}$) is a creation (annihilation) operator for a magnetoferron in band $\nu$ with momentum $\bq$ and ${\rm t}_{\bq, \ell, \nu}$ corresponds to the suitably normalized eigenvectors of Eq. (\ref{eq: eigensystem}). This equation opens the doors to quantum phenomena, such as \cite{Tapia2024, RoldanMolina2015, RoldanMolina2016, RoldanMolina2017, Doornenbal2019}, in the context of quantized magnetoferrons.

Additional symmetry-respecting coupling of the polarization to the strain field will likely open the way to magnetoelastic phenomena such as piezomagnetism\cite{Jaime2017} or the piezospintronic effect\cite{Nunez2014, Ulloa2017, Guo2020}.

The unique properties of magnetoelectric waves in multiferroic materials shall lead to the development of various applications. The propagation and control of the magnetoferrons entail the possibility of controlling spin ($\vec{j}_S$)\cite{Harms2022}, polarization ($\vec{j}_P$)\cite{Bauer2023}, and heat ($\vec{j}_q$)\cite{Wooten2023} currents with great control, especially near the critical region.
The first idea that comes to mind is a magnetoelectric sensor that detects weak magnetic fields. These sensors might utilize the interferometry of magnetoferrons and their dependence on electric and magnetic properties in multiferroic materials to detect changes in magnetic fields by measuring the induced electric polarization. These sensors might find applications in fields as different as biomagnetic sensing, geophysical exploration, and non-destructive testing.
The second is a neuromorphic memory device that stores data in both the electric polarization and magnetization states of multiferroics. This could be implemented by the design of magnetoferron-based memristors\cite{Tetzlaff2013}.
Information is customarily written and read using electric and magnetic fields. The emergence of magnetoferrons and the coupling between them and these fields in multiferroic materials allows efficient data manipulation. This opens the way for data storage solutions that offer faster access times and lower power consumption than traditional memory devices.
Third, a spintronic device that uses magnetoelectric coupling to control the spin current. The device exploits the magnetoelectric effect to control the spin orientation of electrons, which is crucial for developing advanced spintronics applications: next-generation transistors, magnetic random access memory (MRAM), and other spin-based electronic devices.
Finally, magnetoferrons can be used to envision a microwave signal processor utilizing magnetoelectric waves for frequency modulation and filtering. Magnetoelectric materials are used to create components such as filters and modulators that operate at microwave frequencies, taking advantage of their unique dispersion properties in telecommunications, radar systems, and wireless communication technologies.

\sectioncustom{Acknowledgements}
Funding is acknowledged from Fondecyt Regular 1230515 and  DICYT regular 042431AP. 
 G.S.  thanks the financial support provided by ANID Subdirección de Capital Humano/Doctorado, Chile Nacional/2022-21222167. M. A. C. acknowledges Proyecto ANID Fondecyt de Postdoctorado 3240112. 

\end{document}